# High-Throughput NEB for Li-Ion Conductor Discovery via Fine-Tuned CHGNet Potential

Jingchen Lian[a,b], Xiao Fu[a,c], Xuhe Gong[a,d], Ruijuan Xiao[a,b*], Hong Li[a,c]

[a] Institute of Physics, Chinese Academy of Sciences, Beijing, 100190, China.

[b] School of Physical Sciences, University of Chinese Academy of Sciences, Beijing 100049, China

[c] Center of Materials Science and Optoelectronics Engineering, University of Chinese Academy of Sciences, Beijing 100049, China

[d] School of Materials Science and Engineering, Key Laboratory of Aerospace Materials and Performance (Ministry of Education), Beihang University, Beijing 100191, China.

*Corresponding author. Email: rjxiao@iphy.ac.cn

## ABSTRACT

Solid-state electrolytes are essential in the development of all-solid-state batteries. While density functional theory (DFT)-based nudged elastic band (NEB) and ab initio molecular dynamics (AIMD) methods provide fundamental insights on lithium-ion migration barriers and ionic conductivity, their computational costs make large-scale materials exploration challenging. In this study, we developed a high-throughput NEB computational framework integrated with the fine-tuned universal machine learning interatomic potentials (uMLIPs), enabling accelerated prediction of migration barriers based on transition state theory for the efficient discovery of fast-ion conductors. This framework automates the construction of initial/final states and migration paths, mitigating the inaccurate barriers prediction in pre-trained potentials due to the insufficient training data on high-energy states. We employed the fine-tuned CHGNet model into NEB/MD calculations and the dual CHGNet-NEB/MD achieves a balance between computational speed and accuracy, as validated in NASICON-type $Li_{1+x}Al_xTi_{2-x}(PO_4)_3$ (LATP) structures. Through high-throughput screening, we identified orthorhombic Pnma-group structures ($LiMgPO_4$, $LiTiPO_5$, etc.) which can serve as promising frameworks for fast ion conductors. Their aliovalent-doped variants, $Li_{0.5}Mg_{0.5}Al_{0.5}PO_4$ and $Li_{0.5}TiPO_{4.5}F_{0.5}$, were predicted to possess low activation energies, as well as high ionic conductivity of 0.19 mS/cm and 0.024 mS/cm, respectively.

## 1. Introduction

The integration of inorganic solid-state electrolytes (SSEs) in all-solid-state lithium-ion batteries provides a promising solution to enhance the safety performance compared to liquid electrolytes[1–3]. Moreover, SSEs allow the use of lithium metal anodes, which have an extremely high specific capacity and low electrochemical potential[4], beneficial for improving the energy density of batteries.

In developing SSEs with high ionic conductivity, high-throughput screening plays a vital role in materials exploration and design[5]. Computationally, the Nudged Elastic Band (NEB)[6] and ab initio molecular dynamics (AIMD) methods are widely used to calculate the energy barrier of Li-ion migration and extrapolated ionic conductivity in SSEs. However, the significant computational cost of the density functional theory (DFT)-based NEB and AIMD renders them unsuitable for large-scale screening. Our previous work[7] used machining-learning models to learn barrier values from a large number of materials to effectively accelerate the screening of fast ion conductors, however, this strategy exhibits limited accuracy and the subsequent DFT-NEB calculations for candidates are still necessary. Some methods have been proposed to accelerate the DFT-NEB calculations by estimating the minimum energy path (MEP), such as R-NEB[8], GP-NEB[9], ApproxNEB[10], etc., which facilitate the DFT-NEB process by employing algorithms to speed up the convergence of each path. However, due to the structure-dependent nature of ion migration paths, a universal scheme for selecting initial and final states across different structures is still lacking, preventing the high-throughput NEB implementation. The AIMD simulations describe the self-diffusion of lithium ions and involve long-time simulations to derive ionic conductivity statistically. To extrapolate the precise ionic conductivity at room temperature, MD simulations of hundreds of picoseconds are essential to obtain converged Mean Squared Displacement (MSD) curves[11]. Zhu et al.[12] designed a screening procedure for superionic lithium conductors through short AIMD runs (50 ps) at 800 K and 1200 K ($MSD_{800K}$ >5 Å$^2$ and $MSD_{1200K}/MSD_{800K}$ <7), mitigating to some extent the computational demand of lengthy AIMD runs for fast ion conductor discovery. To address these limitations, this work establishes an automated high-throughput NEB screening workflow which systematically exploring the inequivalent migration paths and integrates the machine learning interatomic potentials (MLIPs) to accelerate both the NEB and MD calculations while maintaining high accuracy.

MLIPs can predict energies and forces with near-DFT accuracy while achieving orders-of-magnitude speed improvement compared to DFT[13]. Notable examples include NequIP[14], Deep Potential[15], and TensorMol[16]. However, most MLIPs are limited to specific systems and elements. The development of universal machine learning interatomic potentials (uMLIPs), based on big materials databases like the Materials Project[17] (MP) containing 89 elements, begins to address this challenge.

Famous uMLIPs like CHGNet[18], M3GNet[19], MACE-MP-0[20], are trained on the DFT-relaxed trajectories from MP data. However, a critical challenge identified by Deng et al.[21] is the softening phenomenon of the potential energy surface (PES) in uMLIPs, which arises from the insufficient high-energy configurations in the training set. This issue becomes particularly pronounced when modeling transition states and other non-equilibrium configurations, underscoring the need to construct a comprehensive materials dataset.

In this work, we developed an automated NEB calculation workflow capable of exhaustively sampling all inequivalent hopping pathways in crystal structures. By integrating the fine-tuned CHGNet potential incorporating transition-state DFT training data, we achieve high-throughput and high-accuracy simulations of ionic migration barriers in crystals. The detailed workflow is illustrated in Fig. 1. Firstly, the transition-state configurations were obtained by the pre-trained CHGNet-based high-throughput NEB (HT-NEB) calculations and the training set with DFT calculations of these transition-states were collected to fine-tune the CHGNet potential. The fine-tuned model was then applied in HT-NEB calculations and MD simulations to obtain the barrier values with high accuracy. The HT-NEB workflow enables us to efficiently obtain precise energy barriers for migration paths in crystal materials. As validated, we performed NEB calculations and MD simulations on the well-known fast ion conductor material $Li_{1+x}Al_xTi_{2-x}(PO_4)_3$ (LATP)[22]. Compared to DFT reference values, the fine-tuned model significantly outperformed the pre-trained model in calculating NEB barriers, MD-derived activation energies, and extrapolated room-temperature ionic conductivity. Additionally, it achieves a substantial speed-up over DFT calculations. We further applied the fine-tuned model to the discovery of fast Li-ion conductors and identified several Pnma space group structures as promising frameworks for fast ion conductors. Notably, the candidate materials exhibited a remarkable increase in ionic conductivity after aliovalent ion doping.

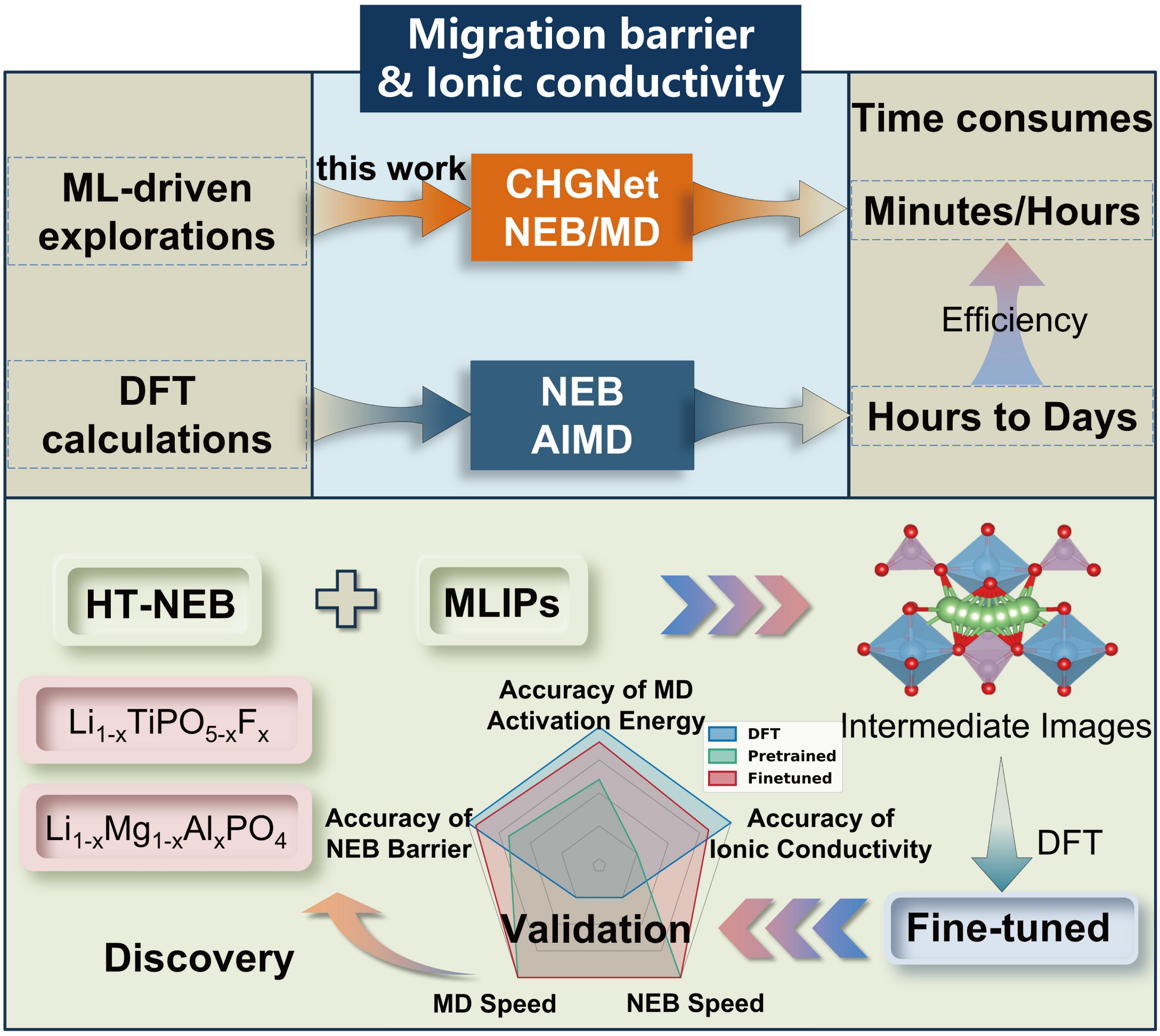

Fig. 1 Schematic of the automated workflow in this study.

## 2. Results and Discussion

### 2.1 Automated High-Throughput NEB workflow

In standard NEB analysis of electrolyte materials, we need to construct supercells to minimize image defect interactions and set initial and final states for ion migration. Subsequently, linear interpolation is applied to generate an initial guess of the minimum energy path (MEP). The climbing-image NEB (CI-NEB) algorithm[23] is then employed in the rigorous convergence calculations to obtain the MEP and energy barrier.

In this work, an automated high-throughput NEB screening workflow is designed to systematically explore all the inequivalent migration paths in each crystal structure. The CIFs obtained from the Materials Project were converted to POSCAR files using Python scripts. To mitigate spurious interactions between migration paths induced by periodic boundary conditions, a supercell with lattice parameters of ~10 Å was constructed for each structure. Site symmetry multiplicities N of Li atoms were directly extracted from CIFs and subsequently employed in the systematic path enumeration process. When the crystal structures exhibit high symmetry with all Li atoms occupying equivalent positions (N = 1), only one migration path requires computation. Whereas complex configurations containing multiple distinct Li sites (designated as site 1, site 2, ..., site N) necessitate consideration of multiple non-equivalent migration channels. An automated vacancy construction method was implemented for generating initial and final states of migration events. The workflow systematically enumerates all inequivalent migration paths by considering each symmetrically distinct $Li^+$ position as an initial state and calculating its hop to the nearest neighbor site for every inequivalent Li type. For instance, if the initial $Li^+$ resides at a crystallographic site $Li_i$ (where $i \in [1, N]$), the process evaluates its migration barriers to all nearest-neighbor sites ($Li_1$, $Li_2$, ..., $Li_N$) associated with other inequivalent Li positions. Consequently, for a structure containing N inequivalent Li sites, the algorithm computes $N^2$ distinct migration pathways. To account for potential energy barrier asymmetry, both forward and reverse hops are explicitly evaluated. This method exhaustively maps all possible $Li^+$ migration channels (denoted as $Li_i \rightarrow Li_j$, where $i, j \in [1, N]$) through combinatorial path enumeration, ensuring complete coverage of inequivalent hops.

An initial guess of the migration path between the initial and final states was first approximated using the Image Dependent Pair Potential (IDPP) method[24], which generates physically realistic atomic trajectories by minimizing interatomic repulsions. Subsequently, NEB calculations were carried out through the Atomic Simulation Environment (ASE)[25] based on CHGNet calculator. The default number of intermediate images in NEB calculations was set to 7 (including endpoints). If the distance between adjacent images exceeds 1 Å, additional intermediate images were automatically inserted to maintain proper connectivity between neighboring states. Using the NEB tool in ASE, we efficiently obtain converged MEP pathways with MLIPs through automated computation.

Take the layered compound $Li_2MnO_3$ as an example. As shown in Fig. 2, the lattice contains three distinct Li sites, in which Li1 and Li2 are located in lithium layer while Li3 stay in the transition-metal layer. Previous DFT calculations have revealed relatively low intralayer migration barriers between Li1 and Li2, and slightly higher interlayer migration barriers[26]. The CHGNet-based NEB calculations, listed in Table 1, mapped all possible migration paths and correctly reproduced the relative ease of Li migration within the Li plane (between 4h and 2c) versus the higher-barrier hops between neighboring Li and $LiMn_2$ layers (between 2b to others).

Table 1 Systematically barrier predictions through NEB method for all the inequivalent Li

migration paths in $Li_2MnO_3$ (mp-18988) by pre-trained and fine-tuned CHGNet potential. Barriers from the initial state (IS) to the final state (FS) are listed. For comparison, the DFT-calculated values from Reference 26 are included.

| IS→FS \ Methods | Pre-trained | Fine-tuned | DFT values[26] |
|---|---|---|---|
| **$Li_1$ (4h)→$Li_1$ (4h)** | 0.50 | 0.70 | 0.74 |
| **$Li_1$ (4h)→$Li_2$ (2c)** | 0.41 | 0.57 | 0.54 |
| **$Li_1$ (4h)→$Li_3$ (2b)** | 0.38 | 0.56 | 0.59 |
| **$Li_2$ (2c)→$Li_1$ (4h)** | 0.42 | 0.58 | 0.61 |
| **$Li_2$ (2c)→$Li_2$ (2c)** | 1.86 | 2.53 | - |
| **$Li_2$ (2c)→$Li_3$ (2b)** | 0.39 | 0.55 | 0.51 |
| **$Li_3$ (2b)→$Li_1$ (4h)** | 0.49 | 0.67 | 0.80 |
| **$Li_3$ (2b)→$Li_2$ (2c)** | 0.48 | 0.64 | 0.73 |
| **$Li_3$ (2b)→$Li_3$ (2b)** | 6.31 | 6.40 | - |

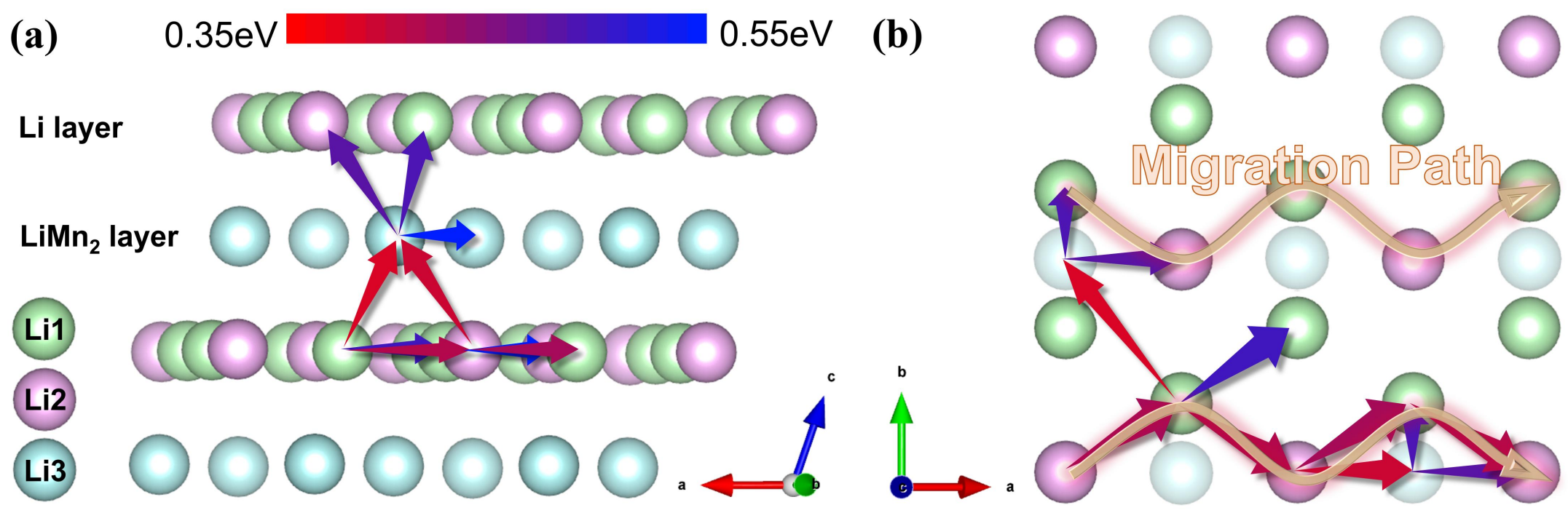

Fig. 2 Different views of Li sites and all enumerated migration paths in $Li_2MnO_3$ (mp-18988). Predicted migration barriers by pre-trained CHGNet potential are visualized via colored arrows (red hues denote lower barriers). A low-barrier conduction network exists along $Li_1 \rightarrow Li_2 \rightarrow Li_1$. While the barriers from $Li_1/Li_2$ to $Li_3$ are relatively low, the migration from $Li_3$ to other Li ions is hard, due to the lower site energy of $Li_3$.

## 2.2 Fine-tuned CHGNet Potential

### 2.2.1 Dataset selection

We systematically constructed the dataset for solid electrolyte discovery based on multiple criteria from the Materials Project[17] database. Candidate selection proceeded according to the following four key criteria: (1) We chose Li-containing quaternary compounds, where quaternary systems were prioritized to encompass polyanionic frameworks and mixed-anion systems, both critical for Li conduction. (2) We narrowed compounds containing only elements commonly found in lithium battery materials as illustrated in Fig. 3 (a). (3) Enforcing maximum oxidation states ensured all candidates are intrinsically stable against further oxidation. (4) We filtered out low-symmetry structures with distinct Li sites larger than 3 and number of atoms in supercell larger than 300 to decrease the total computational cost of DFT. These structures were processed by the HT-NEB workflow introduced in the last section, which was performed with the pre-trained CHGNet potential for these candidates and effectively created transition states for all the inequivalent pathways in each structure.

The dataset containing 3,115 transition-state configurations was generated. DFT static calculations were performed on each configuration to obtain the energies, forces, and stresses required for fine-tuning the CHGNet potential. Fig. 3 (b) illustrates the elemental distribution in the dataset, where the *x*-axis lists elements and the y-axis represents the percentage of materials containing each element. From this dataset, 2,784 configurations were randomly selected as the training set, which was partitioned into training (80%), validation (10%), and test (10%) subsets for fine-tuning the CHGNet potential. The remaining 331 configurations constituted a separate test set.

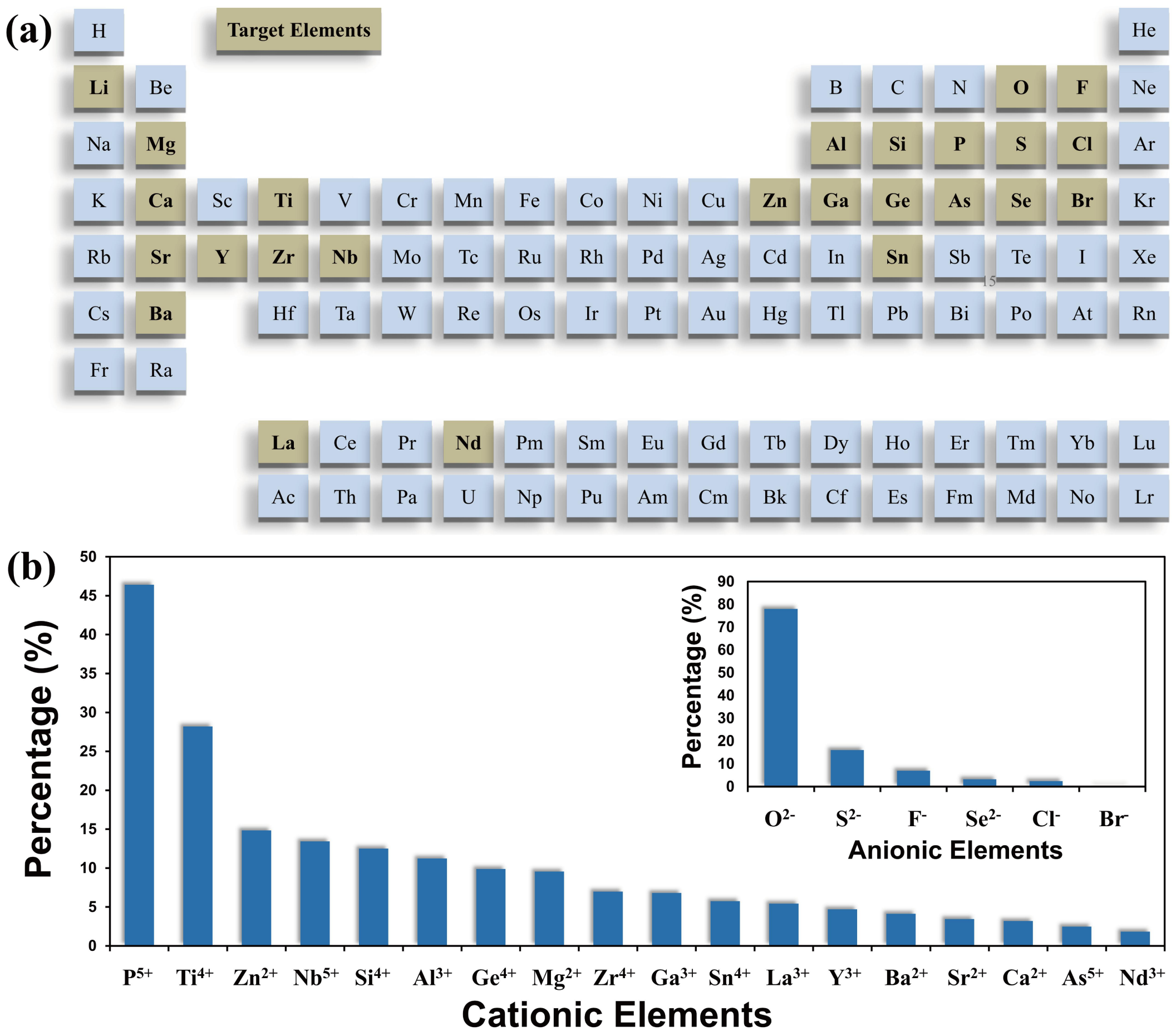

Fig. 3 (a) Target elements filtered through the screening workflow. (b) Cationic and anionic elements and related percentages to the whole number of target compounds in the dataset.

### 2.2.2 Fine-tuned Potential

The fine-tuning of CHGNet potential is performed with the DFT transition state datasets mentioned in the subsection 2.2.1. The specific parameters used for the fine-tuning process are described in the method part in detail. The fine-tuned CHGNet potential achieved much better performance with the mean absolute errors (MAE) of 2 meV/atom for energy, 13 meV/Å for force, and 13 mGPa for stress. The comparison between the pre-trained and fine-tuned model is illustrated in Fig.S1.

Besides the improvement of the model accuracy, the fine-tuned model also demonstrates enhanced precision in energy barrier predictions. Fig. 4 (a) and (b) compare the migration barriers predicted by both pre-trained and fine-tuned CHGNet models against DFT reference values for the training and separate test sets. Due to the computational cost of DFT-based NEB calculations, the DFT barrier references were constructed by computing single point energies at CHGNet-predicted transition states, shown as x-axis, and the y-axis shows the barriers predicted by NEB calculations with the two CHGNet models respectively. The fine-tuned model reduced the MAE of barriers prediction from 0.24 eV to 0.07 eV on the training set and from 0.23 eV to 0.09 eV on the test set. Compared to the pre-trained model, the fine-tuned model improved the $R^2$ value from 0.97 to 0.99 on the training set and from 0.94 to 0.98 on the test set. These results demonstrate that the fine-tuned model achieves significantly better agreement with DFT predictions across both datasets.

To further demonstrate the general improvement of the fine-tuned model in mitigating potential energy surface softening, we analyzed the migration paths with 7 images selected from both the training and test sets. The energy error for each image was statistically represented using a violin plot, as shown in Fig. 4 (c) and (d). Here, image 0 corresponds to the initial state, where the DFT and CHGNet energies are aligned, while image 7 represents the final state after lithium migration. We observed that as the image index approaches the midpoint where Li is near the energy maximum, both CHGNet models tend to underestimate the energy barriers relative to DFT values. However, the fine-tuned model exhibits lower median energy errors, reduced interquartile range (IQR), and smaller extrema (details shown in Table S1 and Table S2). These improvements indicate that the fine-tuned model significantly mitigates the softening effect of the potential energy surface, making it more suitable for accurately describing high energy-state structures in NEB and MD simulations.

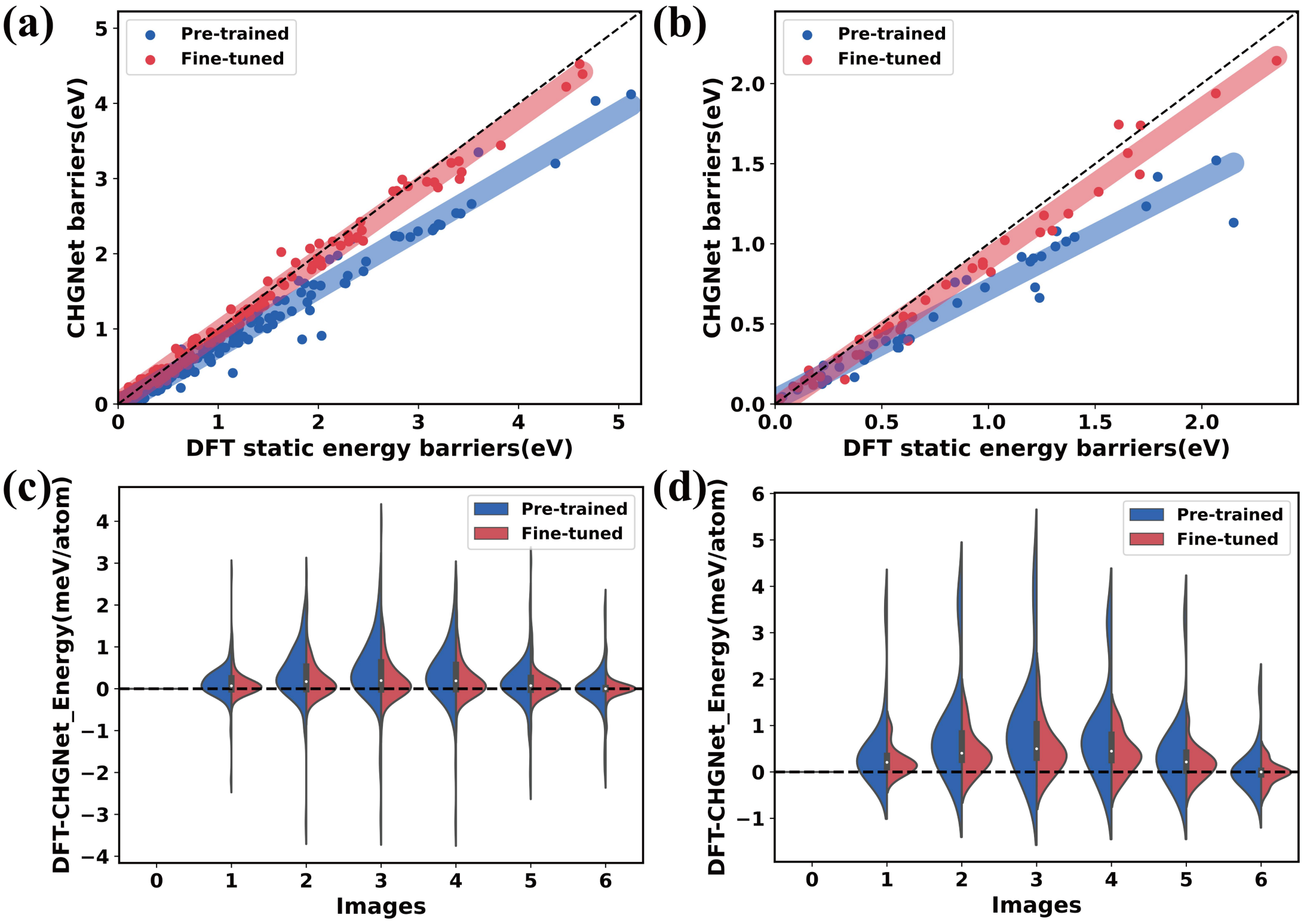

Fig. 4 Systematic validation of CHGNet-based NEB barrier prediction accuracy. Comparing DFT static energy barriers (x-axis) and CHGNet-NEB predictions (y-axis) for (a) training set (MAE reduced from 0.24 eV to 0.07 eV, $R^2$ improved from 0.97 to 0.99) and (b) test set (MAE

reduced from 0.23 eV to 0.09 eV, $R^2$ improved from 0.94 to 0.98). Violin plots quantifying energy errors for 7-image pathways in (c) training set (mean error for energy decreases from 0.45 meV/atom to 0.25 meV/atom for configurations at image 3) and (d) test set (mean error for energy decreases from 1.01 meV/atom to 0.49 meV/atom for configurations at image 3), with boxplots showing interquartile ranges (IQR). Dashed lines indicate 0 meV error.

## 2.3 **High-throughput MD workflow**

To determine the activation energies ( $E_a$ ) and the room-temperature ionic conductivity within the high-throughput framework, the CHGNet based-MD workflow was also established. For comparison, AIMD simulations with identical ensemble, timestep, and simulation time parameters were performed as reference. Due to substantial statistical variations between independent MD runs[11], it is necessary to perform multiple long-duration MD simulations. Therefore, for each temperature, we typically conducted three MD simulations, each lasting 200 ps. To improve statistical convergence, we divided these long trajectories into 12 non-overlapping 50 ps segments. We then averaged the resulting MSD curves and determined the diffusion coefficient at each temperature from their slopes. To mitigate nonlinear artifacts at the endpoints of $\Delta t$, we restricted the linear fits in the 20%-80% range of $\Delta t$. Furthermore, to address potential changes of activation energy at high temperatures due to phase transitions or alterations in migration mechanisms[27], our approach involves fitting the data points at lower temperatures (meanwhile ensuring the linearity of MSD curves by extending the simulation time) while assuming constant $E_a$, allowing us to extrapolate the room temperature conductivity by the Nernst-Einstein equation:

$$\sigma(T) = \frac{ne^2z^2}{k_BT}D(T) \qquad (1)$$

where $n$, $z$ represents the volume density ($cm^{-3}$) and the charge of diffusing species (+1 for lithium ions), $D(T)$ represents the diffusion coefficients at a given temperature.

While AIMD is limited by its high computational cost, the efficiency of CHGNet potentials enables long-timescale MD simulations, particularly crucial for systems with rare migration events at low temperatures, to achieve well-converged diffusion statistics. Meanwhile, when extrapolating the room temperature conductivity by the diffusion coefficients at multiple temperatures, CHGNet-based MD can offer more reliable $D(T)$ data points to enhance the accuracy of Arrhenius fitting. Moreover, for most studied systems, CHGNet-based MD can directly simulate ionic conductivity at target temperatures, thereby eliminating the need for extrapolation procedures.

## 2.4 **Validation for the Fine-tuned CHGNet**

In this section, we take the specific examples to verify the accuracy of the fine-tuned CHGNet model in both NEB calculations and MD simulations. The energy barriers for all the inequivalent Li migration paths for $Li_2MnO_3$ have been listed in Table 1. The energy barriers predicted by the pre-trained model are obviously lower than DFT values calculated by previous work[26], with an MAE of 0.21 eV. Although Mn element is not included in the fine-tuning training set, the more accurate description of Li-O interactions in the fine-tuned model improves the barriers prediction with a lower MAE of 0.056 eV.

Besides the layered $Li_2MnO_3$, we also examine the two models on $LiTi_2(PO_4)_3$ (LTP), along with its widely adopted derivative LATP solid electrolyte. DFT-based NEB calculations have revealed that Li ions migrate in a vacancy mechanism with a barrier of about 0.41 eV in the pure $LiTi_2(PO_4)_3$, while the interstitial mechanism with a lower calculated barrier of 0.19 eV occurs in the LATP structure[28]. For a precisely proportioned $Li_{1.5}Al_{0.5}Ti_{1.5}(PO_4)_3$, Wang et al[29] got a 0.23 eV of Li diffusion by the AIMD method. Experimental measurements indicated that a high ionic conductivity of about 1 mS/cm and low activation energy of about 0.28 eV can be achieved in LATP samples synthesized by melt queening method[30], mechanical activation method[31], and sol-gel method[32].

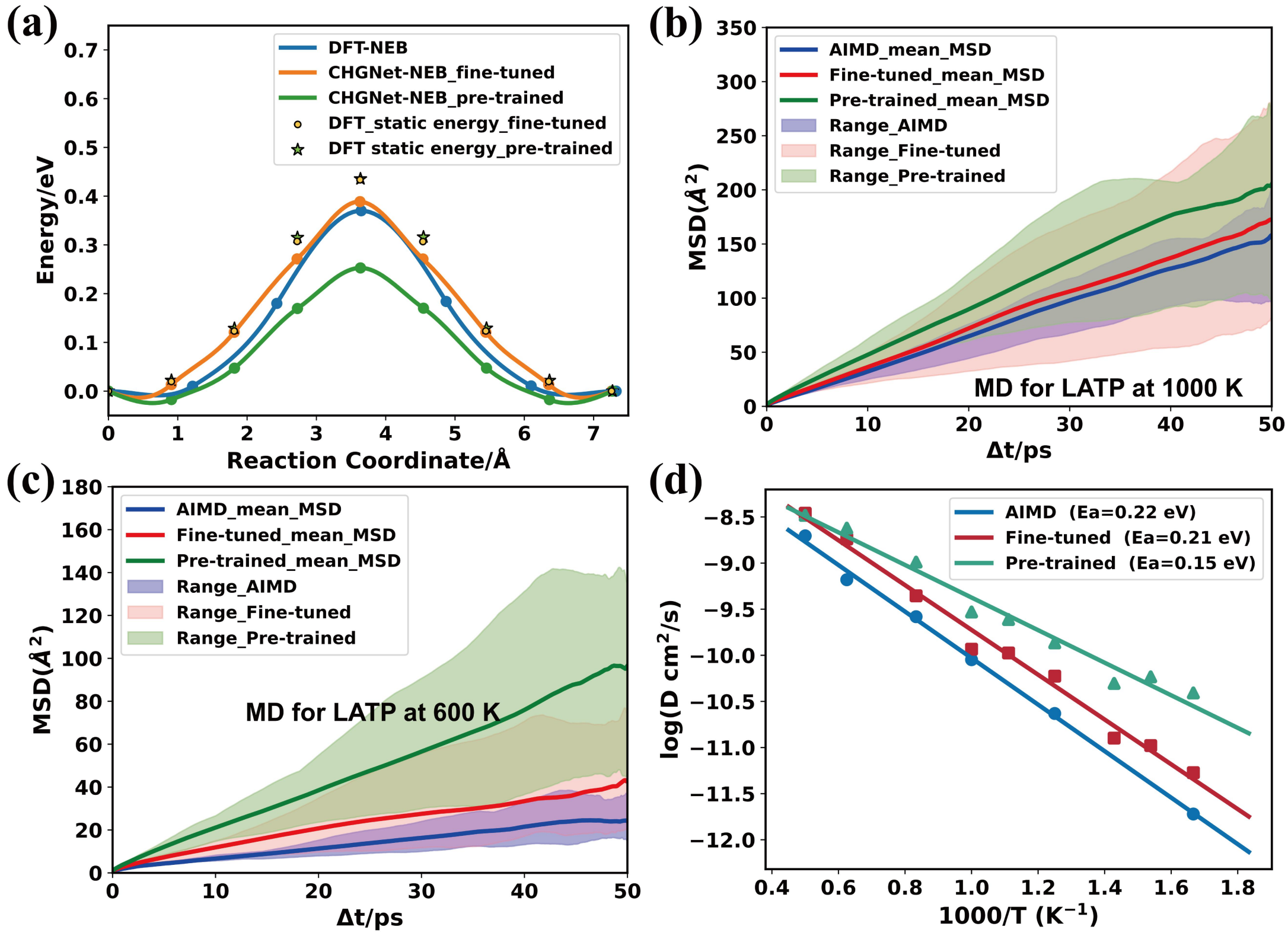

Fig. 5 Comparison of CHGNet with DFT calculations in NASICON-type electrolytes $LiTi_2(PO_4)_3$ and $Li_{1.5}Al_{0.5}Ti_{1.5}(PO_4)_3$: (a) Energy barriers from DFT-NEB and CHGNet-NEB calculations in $LiTi_2(PO_4)_3$; (b)(c) MSD profiles of $Li_{1.5}Al_{0.5}Ti_{1.5}(PO_4)_3$ at (b) 1000 K and (c) 600 K, respectively, comparing AIMD, fine-tuned CHGNet model, and pre-trained CHGNet results; (d) Arrhenius

plot of LATP ionic conductivity showing improved agreement between fine-tuned CHGNet and AIMD.

Fig. 5 (a) presents the NEB results for $LiTi_2(PO_4)_3$. The fine-tuned CHGNet model predicts a barrier of 0.40 eV, which is close to the DFT-NEB result of 0.38 eV, representing an 80% error reduction compared to the pre-trained model's prediction (0.28 eV). For more complex doped systems, such as $Li_{1+x}Al_xTi_{2-x}(PO_4)_3$ (LATP), where aluminum doping introduces interstitial Li ions and the reduced crystal symmetry creates numerous inequivalent migration pathways, making the HT-NEB method impractical for mapping all the energy barriers completely. To address this, MD simulations are more suitable for calculating migration barriers in doped systems. The light-shaded regions in Fig. 5 (b) and (c) represent the dispersion of the MSD curves simulated for LATP. The diffusion coefficient at a given temperature was calculated from the slope of the averaged MSD curve. At 1000 K, the fine-tuned CHGNet model yielded a diffusion coefficient ($5.64\times10^{-5}$ $cm^2/s$) closely aligned with AIMD results ($5.35\times10^{-5}$ $cm^2/s$). The agreement keeps at 600 K with a diffusion coefficient of $1.15\times10^{-5}$ $cm^2/s$ by fine-tuned model versus $8.17\times10^{-6}$ $cm^2/s$ by AIMD. By fitting the diffusion coefficients at various temperatures, the migration barrier and room temperature conductivity can be determined using the Arrhenius equation. The fine-tuned model predicts a migration barrier of 0.21 eV, which is close to the AIMD result of 0.22 eV. The predicted lithium-ion conductivity at 300 K is 8.6 mS/cm, which is close to the AIMD result of 5.1 mS/cm. In contrast, the pre-trained model predicts a conductivity (57 mS/cm) at 300 K, which is an order of magnitude higher. Due to the high efficiency of MD simulations by the fine-tuned CHGNet model, we conducted three separate MD runs at 300 K. The resulting mean MSD curve yielded a room-temperature conductivity of 2.8 mS/cm, which aligns more closely with the experimental value of 1 mS/cm discussed earlier. The fine-tuned CHGNet model not only significantly improves the accuracy of NEB and MD calculations but also achieves a substantial speedup compared to first-principles calculations, as detailed in the Supplementary Information.

## 2.5 Discovery of $Li^+$ fast ion conductors

Although we fine-tuned CHGNet model using only quaternary compounds, its application is not limited to these systems. By accurately capturing the interaction behavior of Li ions, other cations and anions, this fine-tuned CHGNet potential can be extended to ternary, quinary, and other multicomponent systems containing elements presented in the fine-tuning dataset.

Here we continue to use the quaternary structure dataset to demonstrate the discovery of novel ionic conductors. 66 compounds were identified with Li-ion migration barriers lower than 0.5 eV through fine-tuned CHGNet-NEB high-throughput calculations. Table. S3 shows their Materials Project identifiers (mp-id), thermodynamic stability ($E_{hull}$ as energy above hull), and migration barriers.

The screening results identify multiple orthorhombic Pnma space group compounds, including $LiMgPO_4$, $LiMgAsO_4$, $LiTiPO_5$, $LiTiAsO_5$, and $LiSiPO_5$. The Mg-based and Ti-based oxides exhibit a low energy above hull (< 50 meV/atom) and have been experimentally observed, while $LiSiPO_5$ shows a large energy above hull of 109 meV/atom. Further analysis of formation energies reveals that while these configurations exhibit low ionic migration barriers, their high Li-vacancy formation energies intrinsically limit charge carrier formation. This is evidenced by the fined-tuned CHGNet-based MD simulations for defect-free configurations, which yield much higher energy barriers,1.9 eV and 1.54 eV for $LiMgPO_4$ and $LiTiPO_5$ respectively, than NEB predictions. To utilize these low-migration-barrier frameworks, we introduced some lithium vacancies by aliovalent cation doping, including (substituting $Mg^{2+}$ with $Al^{3+}$ in $LiMgPO_4$, and $O^{2-}$ with $F^-$ in $LiTiPO_5$). This doping strategy significantly decreases the barriers, reducing the activation energy to 0.38 eV and 0.52 eV for the two respective configurations. Fitting the diffusion coefficients to the temperature using the Arrhenius relationship, we get room-temperature conductivity of 0.19 mS/cm for $Li_{0.5}Mg_{0.5}Al_{0.5}PO_4$ and 0.024 mS/cm for $Li_{0.5}TiPO_{4.5}F_{0.5}$. Both doped structures maintained reasonable thermodynamic stability with the energy above hull values of 29.7 meV/atom and 0 meV/atom.

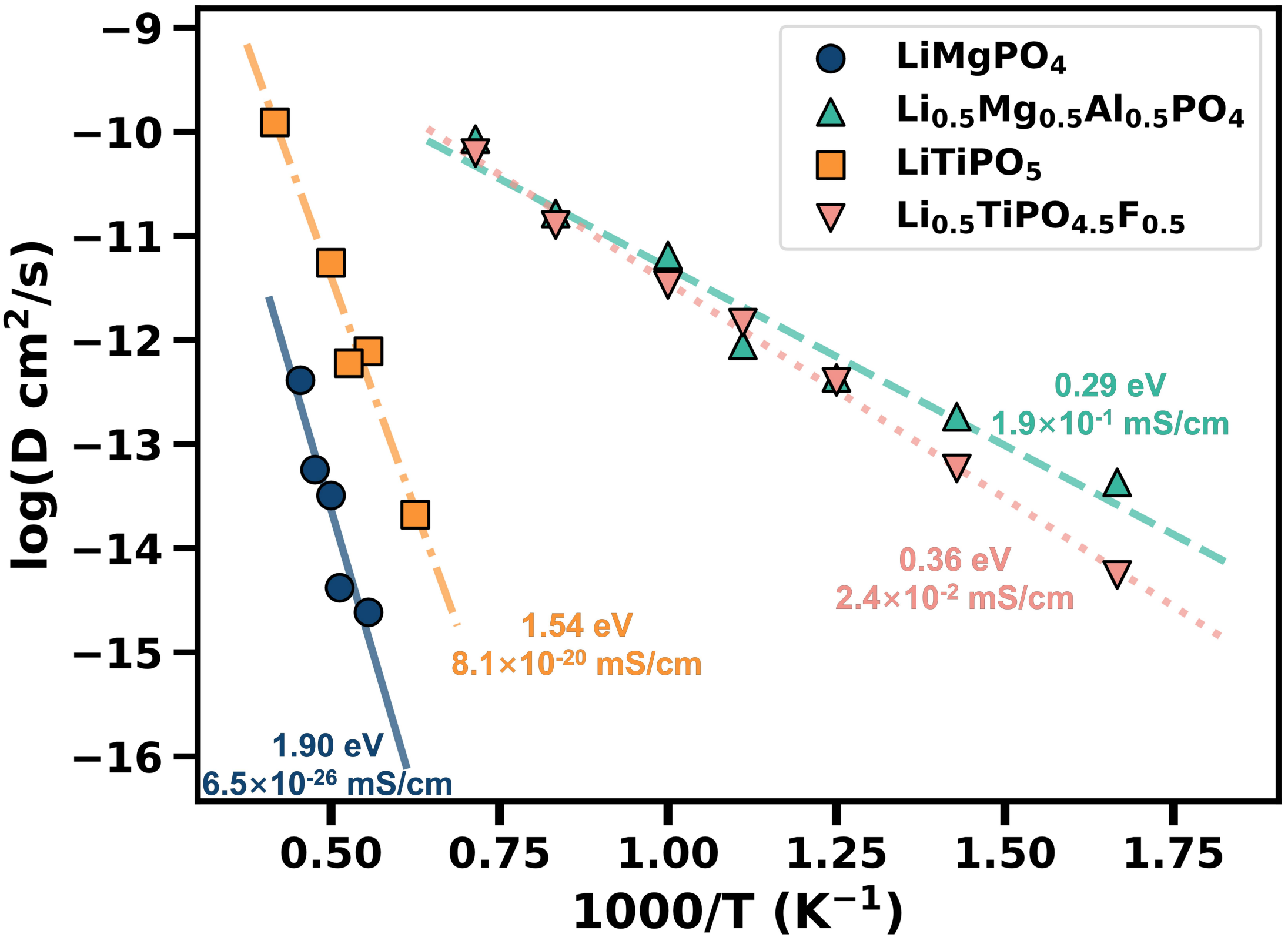

Fig. 6 Arrhenius plot of $LiMgPO_4$, $LiTiPO_5$ and their doped structures simulated by fine-tuned CHGNet-MD.

## 3. Conclusions

In this work, we have developed an automated high-throughput NEB screening workflow integrated with fine-tuned CHGNet model which systematically explores all the inequivalent migration paths in crystal structures and predicts the Li-ion migration barriers in accuracy. The CHGNet-based NEB method demonstrates significant improvements over traditional approaches in three aspects:

(1) Automates the NEB calculation process including cell expansion, initial/final state construction, and IDPP interpolation for path initial guess.

(2) Using CHGNet as the energy and force calculator enables rapid optimization of transition states in NEB calculations to locate saddle points along the pathway. Compared to the pre-trained potential, the fine-tuned CHGNet model mitigates the potential energy surface softening obviously and demonstrates an 80% improvement in energy barrier prediction accuracy.

(3) The high-throughput CHGNet-NEB framework enables the efficient discovery of new superionic conductors. And the CHGNet-MD method provides an efficient approach for studying ionic conductivity in low-symmetry doped structures even at low-temperatures with rare migration events. In comparison to traditional method, the fine-tuned CHGNet HT-NEB/MD simulations achieve a balance between accuracy and efficiency, making it suitable for large-scale screening of low-barrier conductive materials.

Through the application of the workflow, we identified 66 compounds with migration barriers below 0.5 eV, particularly noting that those belonging to the Pnma space group displayed low barriers and high stability, such as $LiMgPO_4$ and $LiTiPO_5$. Based on this structural framework, the doped structures $Li_{0.5}Mg_{0.5}Al_{0.5}PO_4$ and $Li_{0.5}TPO_{4.5}F_{0.5}$ exhibiting high ionic conductivity (0.19 mS/cm and 0.024 mS/cm) were explored. Reasonable thermodynamic stability ($E_{hull}$ of 29.7 meV/atom and 0 meV/atom, respectively) was maintained. This study provides a new strategy for developing novel solid-state electrolytes by using machine-learning interatomic potentials fine-tuned by high-energy structures. This approach enables large-scale conductivity predictions for complex doped structures, facilitating the discovery of next-generation fast-ion conductors.

## 4. Methods

All DFT calculations were performed by Vienna ab initio Simulation Package (VASP)[33] within the projector augmented wave approach (PAW) with the Perdew-Burke-Ernzerhof (PBE)[34] generalized gradient approximation. A uniform Monkhorst-Pack kpoint mesh was generated for each structure such that the spacing between adjacent k-points in reciprocal space did not exceed 0.05 $Å^{-1}$, ensuring consistent sampling density across different unit cell sizes. For AIMD simulations an NVT ensemble of Nose-Hoover thermostat [35] and a timestep of 2 fs were used to accelerate the long-duration simulations. CHGNet molecular dynamics were simulated with pre-trained/fine-tuned CHGNet model through ASE python interface, with an NVT ensemble and a timestep of 1fs.

The training set containing DFT static energy calculations for a total of 2,784 structures was generated by the NEB method using the pre-trained CHGNet model. All images were calculated by VASP using the projector-augmented wave (PAW) method to get single point energy, atomic forces and lattice stress as labels. CHGNet model was fine-tuned by these data with energy, force and stress labels with normalized loss weights (energy: 1, forces: 1, stresses: 0.1) under the mean squared error (MSE) optimization. The dataset was partitioned into training (80%), validation (10%), and test (10%) subsets. RAdam optimizer and initial 0.001 learning rate were used to perform 500 epochs training. The optimized model achieved a good performance with the MAE of 2 meV/atom for energy, 13 meV/Å for force and 13 mGPa for stress (Fig.S2). To further validate the accuracy of the fine-tuned model, we prepared an additional test set comprising 331 configurations. This test set was then used to compare the model's performance on the training set with that on the unseen compositions and structures, allowing us to assess its generalization performance beyond the training data.

## Author Contributions

R.X. designed this work and guided the completion of the method. J.L. constructed the models. All the authors participated in the analysis of the data and discussions of the results, as well as in preparing the paper.

## Conflicts of interest

The authors declare no conflict of interest.

## Acknowledgements

This work was supported by funding from the Strategic Priority Research Program of Chinese Academy of Sciences (grant no. XDB1040302. XDB0500200), and the National Natural Science Foundation of China (grants no. 52172258). The numerical calculations in this study were carried out on both the ORISE Supercomputer, and the National Supercomputer Center in Tianjin.

# High-Throughput NEB for Li-Ion Conductor Discovery via Fine-Tuned CHGNet Potential

Jingchen Lian[a,b], Xiao Fu[a,c], Xuhe Gong[a,d], Ruijuan Xiao[a,b*], Hong Li[a,c]

[a] Institute of Physics, Chinese Academy of Sciences, Beijing, 100190, China.

[b] School of Physical Sciences, University of Chinese Academy of Sciences, Beijing 100049, China

[c] Center of Materials Science and Optoelectronics Engineering, University of Chinese Academy of Sciences, Beijing 100049, China

[d] School of Materials Science and Engineering, Key Laboratory of Aerospace Materials and Performance (Ministry of Education), Beihang University, Beijing 100191, China.

*Corresponding author. Email: rjxiao@iphy.ac.cn

Speed test for CHGNet/DFT-based NEB and MD

In the validation of LATP, CHGNet-based NEB and MD simulations achieved substantial computational speedups compared to first-principles calculations. All the calculations were performed on systems containing approximately 100 atoms using an AMD EPYC platform with an RTX 4090 GPU. For the NEB calculations, the DFT-NEB method requires approximately 10,000 seconds to converge in a criterion of 0.02 eV/Å. In contrast, though operated at a moderate convergence precision of 0.15 eV/Å due to inherent precision limitations, it reduces the computational time to the order of minutes. Similarly, AIMD required about 1.7 h to perform 1000 steps, while the CHGNet-based MD completed $2\times10^5$ steps in 3.2 h, representing a two-order-of-magnitude improvement.

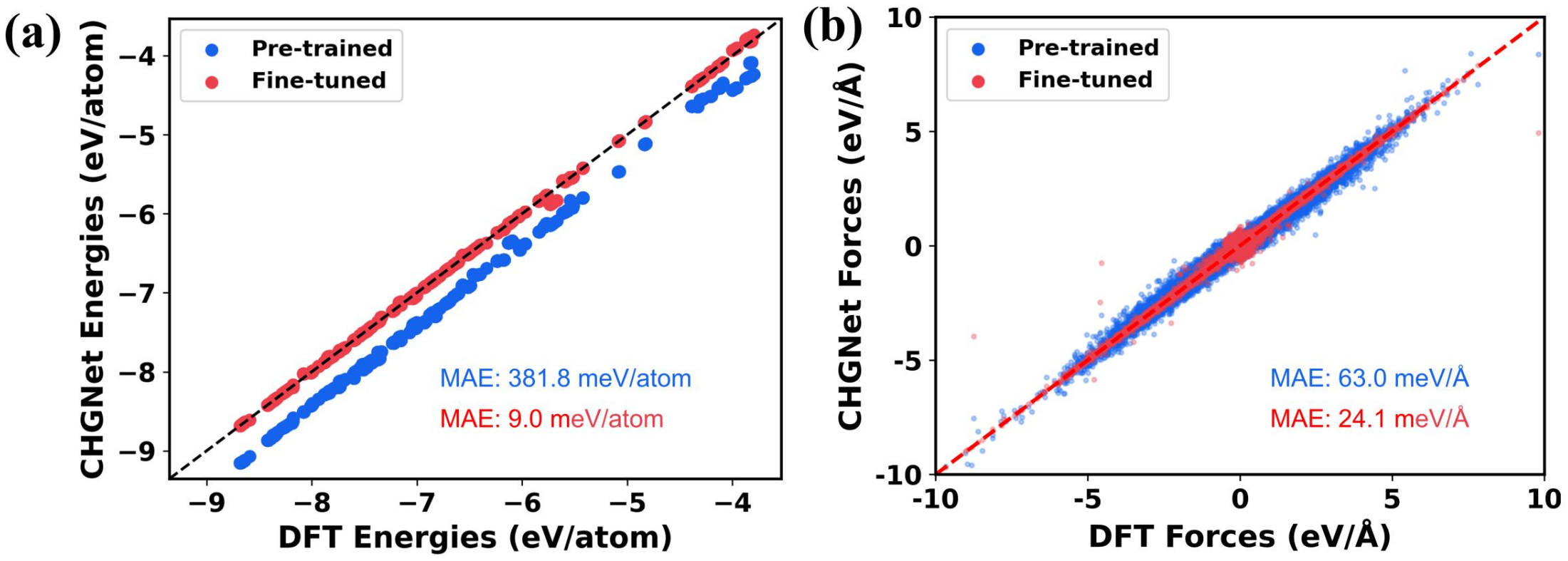

Fig.S1 (a) Energy, (b) force values from the CHGNet pre-trained model and fine-tuned model, in comparison to the DFT reference data in the training set.

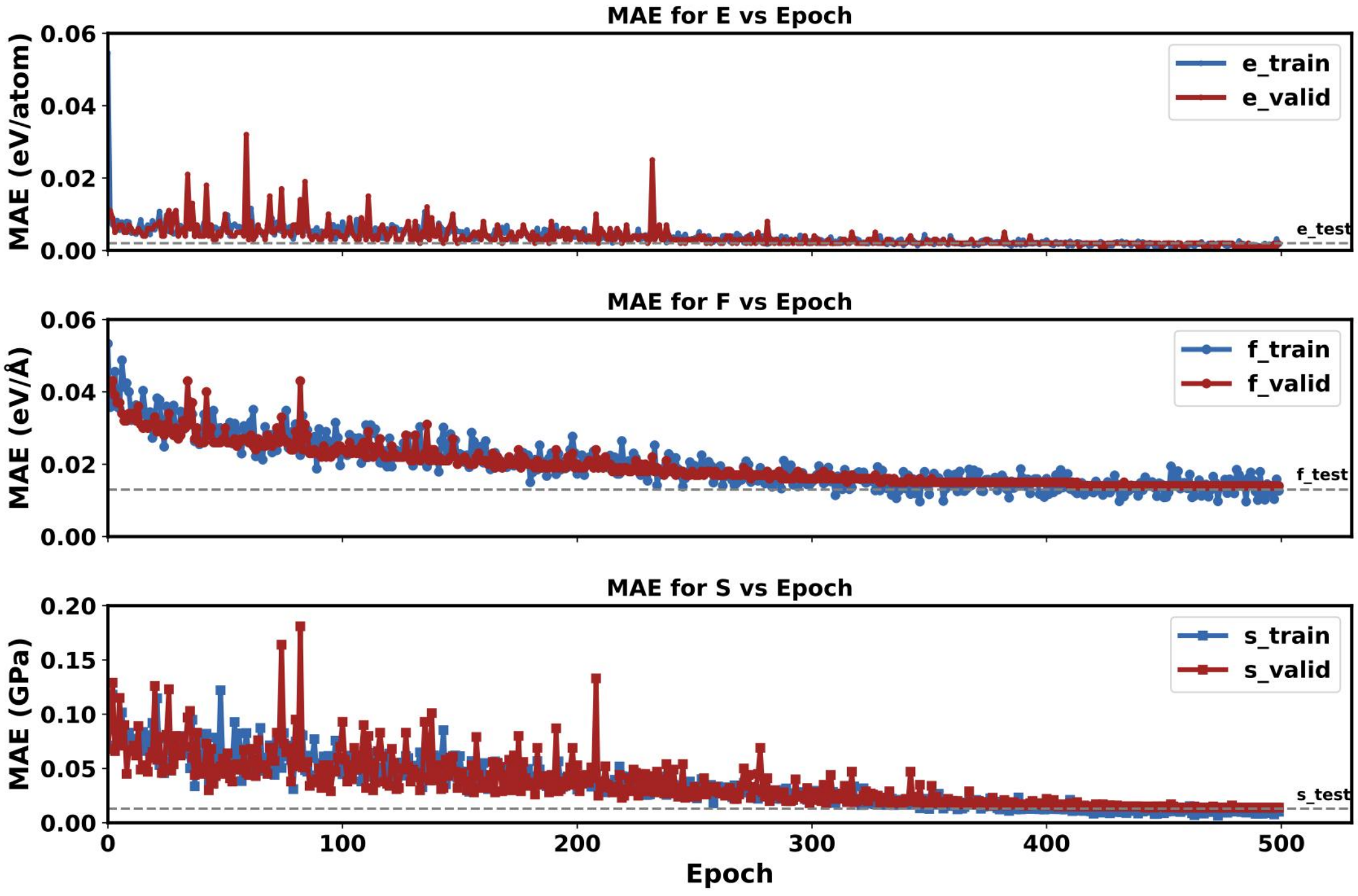

Fig.S2 The mean absolute errors (MAEs) for energy, force and stress in training, validating and testing.

Table S1 Statistic data of violin plot in the training set. The pretrained and finetuned model predicts energies corresponding to each NEB image. The statistic data describes the errors between CHGNet and DFT, includes mean errors (mean), standard deviation (std), median errors (median), first quartile (Q1), third quartile (Q3), interquartile range (IQR), minimum errors (min), maximum errors (max) and sample size (count).

| image | model | mean | std | median | Q1 | Q3 | IQR | min | max | count |
|---|---|---|---|---|---|---|---|---|---|---|
| 0 | pretrained | 0 | 0 | 0 | 0 | 0 | 0 | 0 | 0 | 108 |
| 1 | pretrained | 0.100249 | 0.375643 | 0.09116 | -0.03917 | 0.304715 | 0.343883 | -2.17966 | 1.03256 | 108 |
| 2 | pretrained | 0.35782 | 0.653995 | 0.29812 | 0.054588 | 0.74839 | 0.693803 | -3.19565 | 1.93356 | 108 |
| 3 | pretrained | 0.449662 | 0.790695 | 0.369105 | 0.059798 | 0.808713 | 0.748915 | -3.10669 | 2.6308 | 108 |
| 4 | pretrained | 0.347505 | 0.718014 | 0.26712 | 0.031443 | 0.650275 | 0.618833 | -3.18873 | 2.48042 | 108 |
| 5 | pretrained | 0.102778 | 0.56594 | 0.08588 | -0.0523 | 0.35491 | 0.407208 | -2.19221 | 2.08156 | 108 |
| 6 | pretrained | -0.00933 | 0.530086 | 0 | -0.02858 | 0.041548 | 0.07013 | -1.94838 | 1.94917 | 108 |
| 0 | finetuned | 0 | 0 | 0 | 0 | 0 | 0 | 0 | 0 | 108 |
| 1 | finetuned | 0.158068 | 0.371086 | 0.065615 | -0.01053 | 0.175738 | 0.186265 | -0.26829 | 2.77635 | 108 |
| 2 | finetuned | 0.257751 | 0.503301 | 0.10238 | -0.01532 | 0.377555 | 0.392878 | -0.46926 | 2.73612 | 108 |
| 3 | finetuned | 0.25405 | 0.563287 | 0.10554 | -0.03405 | 0.408973 | 0.443018 | -0.8462 | 3.96786 | 108 |
| 4 | finetuned | 0.244574 | 0.510687 | 0.0759 | -0.01793 | 0.326223 | 0.34415 | -0.89145 | 2.55829 | 108 |
| 5 | finetuned | 0.155927 | 0.426111 | 0.05525 | -0.01458 | 0.197433 | 0.21201 | -0.8866 | 3.04738 | 108 |
| 6 | finetuned | 0.003661 | 0.274901 | 0 | -0.00351 | 0.01652 | 0.020028 | -0.97772 | 0.97793 | 108 |

Table S2 Statistic data of violin plot in the test set.

| image | model | mean | std | median | Q1 | Q3 | IQR | min | max | count |
|---|---|---|---|---|---|---|---|---|---|---|
| 0 | pretrained | 0 | 0 | 0 | 0 | 0 | 0 | 0 | 0 | 19 |
| 1 | pretrained | 0.459919 | 0.803259 | 0.2473 | 0.102675 | 0.435715 | 0.33304 | -0.1206 | 3.46997 | 19 |
| 2 | pretrained | 0.858084 | 1.051335 | 0.70358 | 0.31467 | 0.86455 | 0.54988 | -0.23656 | 3.78189 | 19 |
| 3 | pretrained | 1.014561 | 1.154367 | 0.79786 | 0.425575 | 1.172795 | 0.74722 | -0.29334 | 4.37403 | 19 |
| 4 | pretrained | 0.795043 | 0.968215 | 0.67228 | 0.335735 | 0.89956 | 0.563825 | -0.37629 | 3.31222 | 19 |
| 5 | pretrained | 0.424436 | 0.816573 | 0.29672 | 0.088975 | 0.560405 | 0.47143 | -0.54276 | 3.32849 | 19 |
| 6 | pretrained | 0.06333 | 0.498516 | 0 | -0.03346 | 0.046625 | 0.080083 | -0.6475 | 1.76985 | 19 |
| 0 | finetuned | 0 | 0 | 0 | 0 | 0 | 0 | 0 | 0 | 19 |
| 1 | finetuned | 0.237738 | 0.299436 | 0.15105 | 0.08135 | 0.272725 | 0.191375 | -0.11346 | 0.97489 | 19 |
| 2 | finetuned | 0.424718 | 0.444845 | 0.35293 | 0.16378 | 0.50955 | 0.34577 | -0.17224 | 1.50933 | 19 |
| 3 | finetuned | 0.490718 | 0.538359 | 0.38115 | 0.19616 | 0.620425 | 0.424265 | -0.20806 | 1.96743 | 19 |
| 4 | finetuned | 0.388986 | 0.418 | 0.35312 | 0.155735 | 0.541305 | 0.38557 | -0.29417 | 1.2126 | 19 |
| 5 | finetuned | 0.181706 | 0.334113 | 0.17824 | -0.02475 | 0.31368 | 0.33843 | -0.38507 | 1.11515 | 19 |
| 6 | finetuned | -0.01622 | 0.228343 | 0.00021 | -0.07801 | 0.014175 | 0.092185 | -0.48447 | 0.40597 | 19 |

Table S3 Potential fast ion conductors with at least one migration path below 0.5 eV threshold (Partial listing of 66 totally compounds).

| CIF | Initial_State | component | $E_{hull}$ (meV/atom) | Barrier (eV) | | |
|---|---|---|---|---|---|---|
| | | | | $Li_1$ | $Li_2$ | $Li_3$ |
| mp-10520 | $Li_1$ | $LiNdTiO_4$ | 23 | 0.46 | | |
| mp-1104386 | $Li_1$ | $Li_6PClO_5$ | 7 | 0.15 | | |
| mp-1105323 | $Li_1$ | $LiNdTi_2O_6$ | 34 | | 0.42 | |
| | $Li_2$ | | | 0.43 | | |
| mp-1105479 | $Li_1$ | $LiNdTi_2O_6$ | 10 | 0.47 | | |
| mp-11175 | $Li_1$ | $LiZnPS_4$ | 0 | 0.31 | | |
| mp-11189 | $Li_1$ | $Li_2MgSiO_4$ | 0 | 0.82 | 0.61 | |
| | $Li_2$ | | | 0.38 | 0.68 | |
| mp-1232382 | $Li_1$ | $LiSiPO_5$ | 109 | 0.1 | | |
| mp-13182 | $Li_1$ | $Li_2TiGeO_5$ | 2 | 0.4 | | |
| mp-16691 | $Li_1$ | $SrLi_2Ti_6O_{14}$ | 0 | 0.14 | | |
| mp-22961 | $Li_1$ | $Li_2ZnCl_4$ | 0 | 0.17 | | |
| mp-23416 | $Li_1$ | $Li_2ZnCl_4$ | 2 | 0.15 | | |
| | $Li_2$ | | | 0.31 | | |
| mp-541661 | $Li_1$ | $LiZr_2(PO_4)_3$ | 18 | 0.39 | | |
| mp-554782 | $Li_1$ | $Li_2Ga_2GeS_6$ | 35 | 0.23 | | |
| mp-558083 | $Li_1$ | $BaLi_2Ti_6O_{14}$ | 0 | 0.11 | | |
| mp-559441 | $Li_1$ | $LiTiPO_5$ | 13 | 0.28 | | |
| mp-560058 | $Li_1$ | $Li_9Ga_3P_8O_{29}$ | 13 | 0.05 | 1.79 | 0.01 |
| | $Li_2$ | | | 2.9 | 1.03 | 2.16 |
| | $Li_3$ | | | 0.08 | 1.28 | 0.63 |
| mp-560209 | $Li_1$ | $Li_9Al_3P_8O_{29}$ | 11 | 0.02 | 1.9 | 0.02 |
| | $Li_2$ | | | 2.83 | 1.12 | 2.06 |
| | $Li_3$ | | | 0.02 | 2.13 | 2.21 |
| mp-6113 | $Li_1$ | $LiTiAsO_5$ | 0 | 0.26 | | |
| mp-6332 | $Li_1$ | $Li_2TiSiO_5$ | 0 | 0.39 | | |
| mp-6521 | $Li_1$ | $LiLaTiO_4$ | 29 | 0.4 | | |
| mp-6668 | $Li_1$ | $LiTiPO_5$ | 0 | 0.26 | | |
| mp-774752 | $Li_1$ | $Li_2Ti_3MgO_8$ | 0 | 0.44 | | |
| mp-8870 | $Li_1$ | $LiMgAsO_4$ | 7 | 0.12 | | |
| mp-9406 | $Li_1$ | $Li_2La_2Ti_3O_{10}$ | 20 | 0.39 | | |
| mp-9625 | $Li_1$ | $LiMgPO_4$ | 0 | 0.27 | | |